# Ultra-high-speed high-resolution laser lithography for lithium niobate integrated photonics

Jinming Chen[a], Zhaoxiang Liu[a], Lvbin Song[a], Chao Sun[a], Guanhua Wang[a], and Ya Cheng[a,b,c,*]

[a]The Extreme Optoelectromechanics Laboratory (XXL), School of Physics and Electronic Science, East China Normal University, Shanghai 200241, China

[b]State Key Laboratory of High Field Laser Physics and CAS Center for Excellence in Ultra-intense Laser Science, Shanghai Institute of Optics and Fine Mechanics (SIOM), Chinese Academy of Sciences (CAS), Shanghai 201800, China

[c]State Key Laboratory of Precision Spectroscopy, East China Normal University, Shanghai 200062, China

**ABSTRACT**

Photolithography assisted chemo-mechanical etching (PLACE), a technique specifically developed for fabricating high-quality large-scale photonic integrated circuits (PICs) on thin-film lithium niobate (TFLN), has enabled fabrication of a series of building blocks of PICs ranging from high-quality (high-Q) microresonators and low-loss waveguides to electro-optically (EO) tunable lasers and waveguide amplifiers. Aiming at high-throughput manufacturing of the PIC devices and systems, we have developed an ultra-high-speed high-resolution laser lithography fabrication system employing a high-repetition-rate femtosecond laser and a high-speed polygon laser scanner, by which a lithography efficiency of 4.8 cm$^2$/h has been achieved at a fabrication resolution of 200 nm. We demonstrate wafer-scale fabrication of TFLN-based photonic structures, optical phase masks as well as color printing.

## 1. INTRODUCTION

The combination of advanced materials with high optical performance and continuous improvement of micro/nano fabrication technology has opened a new chapter for integrated photonics. Recently, thin-film lithium niobate (TFLN) has attracted tremendous attention as the material platform for the next generation photonic integrated circuits (PICs)[1-8], mainly owing to its wide transparency window from UV to mid-IR, moderately high refractive index that enables dense photonic integration while maintaining a suitable mode-size in the single-mode LN ridge waveguide, and large electro-optic (EO) as well as nonlinear optical coefficients which are critical for high-speed EO tuning and high-efficiency wavelength conversion applications[9-11]. Notably, a broad range of high-performance PIC devices have been demonstrated using the photolithography assisted chemo-mechanical etching (PLACE) technique[12-20] which is developed specifically for the mass-production of large-scale PICs. As some examples, we have fabricated ultra-high quality (Q~1.23×10$^8$) optical microresonators[17,18] and optical delay lines of meter-scale lengths[16] using the PLACE technique. The PLACE fabrication technique employs laser direct writing for mask patterning which breaks through the size-limit of exposure area and chemo-mechanical polish (CMP) which gives rise to sub-nanometer surface roughness on the sidewall of the fabricated waveguides to efficiently suppress the scattering loss of the propagating light. Meanwhile, it is a well-developed technology that the LN crystal can be used as a host material for doping rare earth ions. Using the doped TFLN, we have demonstrated electro-optically tunable microlaser, single-frequency ultra-narrow linewidth microdisk laser, and waveguide amplifier on Er3+ doped TFLN[21-24].

Owing to its direct-writing nature, the fabrication efficiency of PLACE intrinsically depends on the scan speed of the tightly focused femtosecond laser. In our previous scheme, the direct writing is realized by keeping the laser focal spot stationary while translating the sample in 3D (XYZ) space using a high-precision XYZ motion stage. The highest scan speed achieved in our previous experiment is 80 cm/s with the sophisticated XY stage, whilst the stage much be frequently accelerated or deaccelerated to enable the writing of complex mask patterns. When the motion stage undergoes the acceleration and deacceleration, the laser writing process must slow down for maintaining the high accuracy in positioning of the motion stage, which in turn reduce the fabrication efficiency. To solve the problem, we recently develop an ultra-high-speed high-resolution femtosecond photolithography system which combines a polygon scanner with the high-precision XYZ motion stage. In this manner, the acceleration and deacceleration of the motion stage are eliminated, and the scan speed can reach 2 m/s which is much higher than the maximum speed of the high-precision motion stages, allowing for achieving an ultra-high production efficiency of 4.8-cm$^2$/h at 200-nm fabrication resolution. We demonstrate fabrication of large-scale photonic structures and devices including TFLN-based true optical delay lines, wafer-scale optical phase plate (OPP) as well as wafer-scale laser color printing.

## 2. THE PLACE FABRICATION TECHNIQUE

To begin with, we introduce the fabrication procedures of PLACE technique, which are schematically illustrated in Fig. 1. First, a chromium (Cr) metal film deposited on TFLN is patterned into pre-designed mask shape using femtosecond laser ablation. Subsequently, the TFLN covered with Cr mask undergoes a CMP process, by which the Cr mask pattern is transferred to TFLN as the Cr mask can protect the underneath LN from the CMP process for the high hardness of Cr. At last, the residual Cr mask is removed, followed by an additional CMP process for further smoothing the upper edge of waveguides. This technique allows for the expansion of the footprint of PICs almost without physical limit as the fabrication resolution is only determined by the focal spot size which is irrelevant with the field of view of the objective lens used for focusing the femtosecond laser beam. In addition, the photonic structures fabricated by CMP process can have a surface roughness well below 1 nm, rendering the propagation loss approaching the absorption limit of LN.

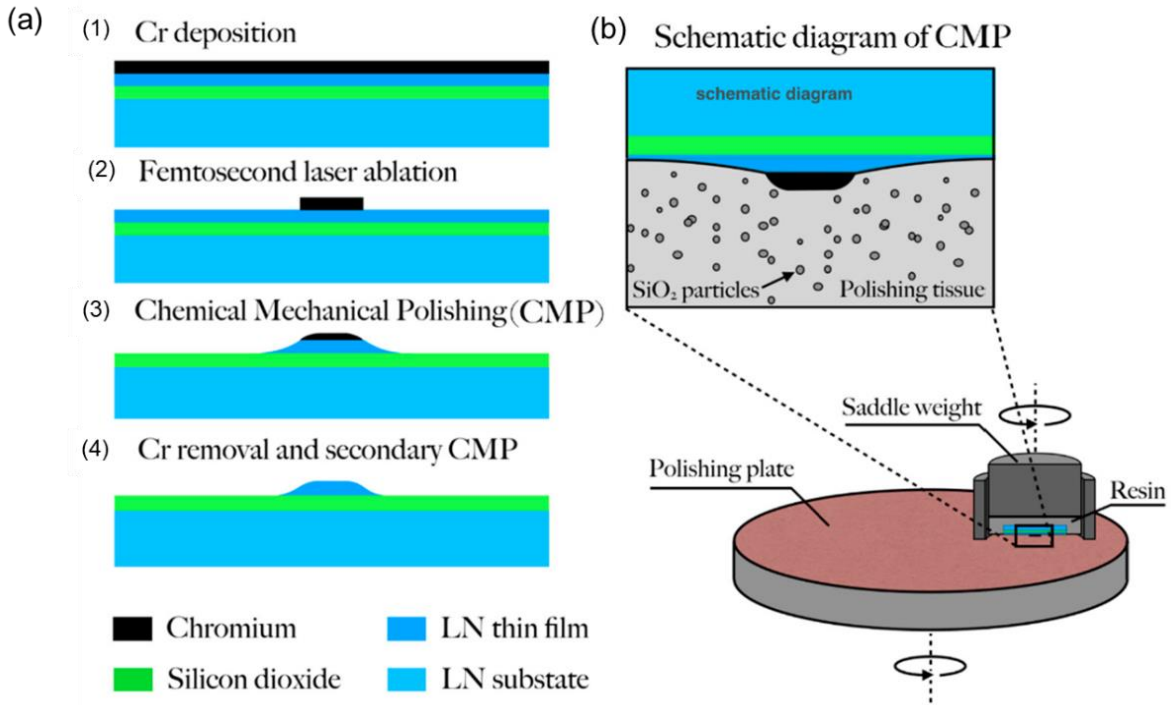

Figure 1. Schematic diagram of (a) PLACE technique processing flows. (b) CMP process.

To showcase the high-performance of the PLACE technique, we fabricated TFLN microdisk[17] and microring[18] resonators, as presented in Fig. 2(a)-(b), respectively, showing an ultra-smooth surface with an average surface roughness below 0.5 nm. The measured transmission spectrum of the microdisk is shown in Fig. 2(c), indicating an intrinsic Q factor of $1.2\times10^8$. Fig. 2(d) presents the Q-factor measured with the photon-ring method which was carried out by repeatedly scanning the laser wavelength with a high-speed Mach-Zehnder modulator into the targeted resonance. The retrieved photon lifetime of 64.3 ns indicates a Q-factor of $1.23\times10^8$, which is in good agreement with the Q-factor in Fig. 2(c). Since the CMP technique does not involve any ion beam etching (IBE) process, surface smoothness beyond that allowed by IBE can be readily achieved. Fig. 2(e) shows a LN waveguide amplifier fabricated on the Er[3+] doped TFLN wafer using the PLACE technique[24]. Thanks to the low propagation loss and long waveguide length of 10 cm, a small signal gain of above 20 dB is achieved at the signal wavelength around 1532 nm when pumped by a diode laser at ~980 nm.

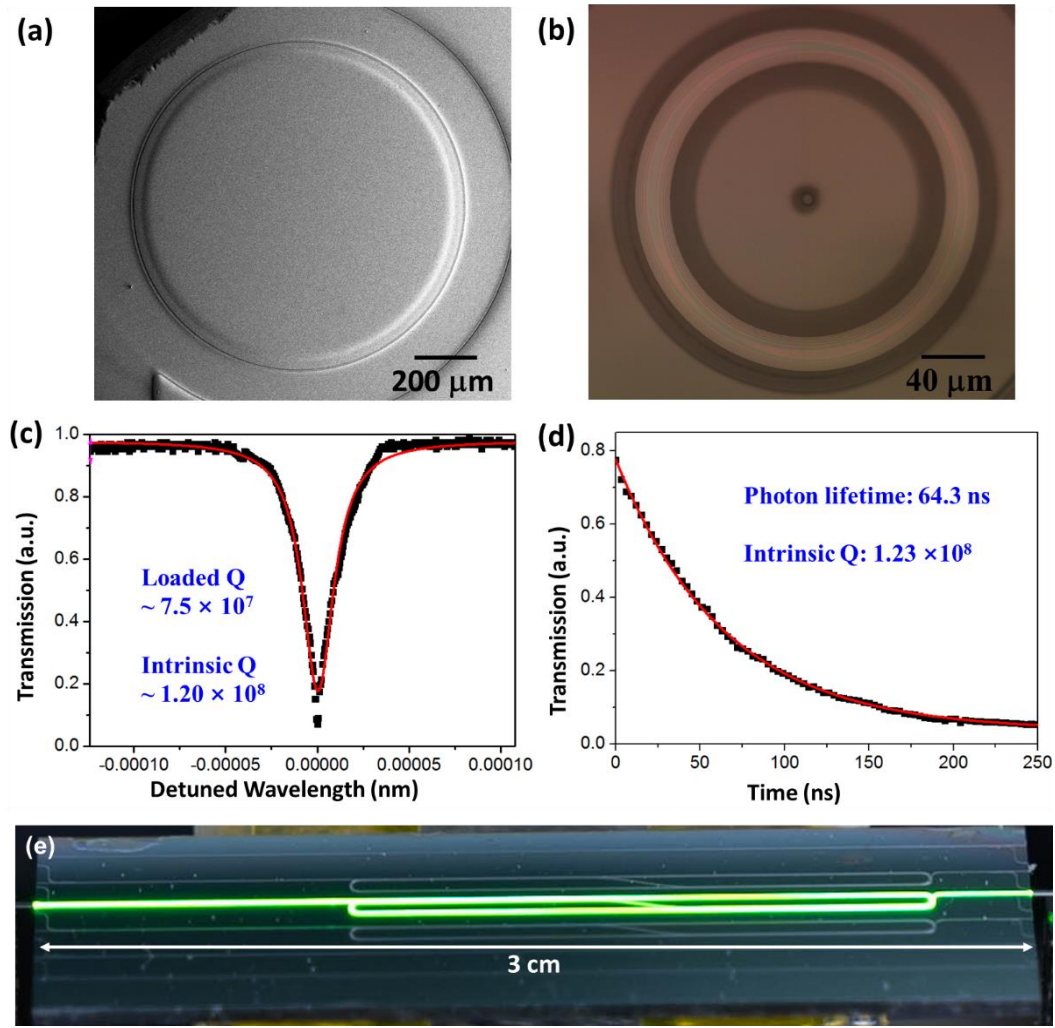

Figure 2. (a) SEM of the microdisk resonator. (b) Optical microscope image of the microring. (c) The transmission spectrum. (d) Photon lifetime measurement at 1551.52 nm wavelength. (e) Digital camera photograph of the excited Er3+-TFLN waveguide chip.

## 3. HIGH-SPEED FEMTOSECOND PHOTOLITHOGRAPHY SYSTEM

The PLACE fabrication technique uses the high-speed femtosecond laser direct writing system for the task of mask patterning on a Cr thin film. The schematic of the newly developed polygon-scanner-based femtosecond lithography system is illustrated in Fig. 3(a), in which a fast-rotating mirror reflects the laser pulses into an objective lens to form a tiny focal spot rapidly moving in the horizontal direction on the top surface of substrate. Meanwhile, the scanning along the vertical direction is realized using an XY motion stage operated at a slower speed to achieve synchronized 2D scan in the XY plane. To form the pre-designed patterns on the Cr film, a home-built controller is used to switch the laser pulses on/off very quickly during the beam scanning. Figure 3(b) shows the photograph of the home-built system which is mainly composed of a high-repetition-rate femtosecond laser system, an XYZ motion stage, and the polygon scanner. In addition, the polygon scanner can be set at a high scan rate of 3000 line/s. By synchronizing the laser pulses, the rotating scanner and the motion stage, we have achieved infinite field of vision (IFOV) processing, i.e. the area on the substrate to be patterned into the mask is no longer limited by the field of view of the objective lens but only limited by the motion range of the XY stage. In our experiment, the motion range of the XY stage is 200 mm×200 mm. For achieving the high-resolution fabrication performance, a high numerical aperture (NA) objective lens is used to carry out the lithography process, resulting in a processing velocity of ~2 m/s and a fabrication efficiency of ~4.8 cm2/h at a fabrication resolution of ~200 nm. In this study, the fabrication resolution is determined by the width of the laser ablation trench formed on the 200 nm-thick Cr film. Surface flatness error is solved by a home-built real-time focusing module. Using the home-built ultra-high-speed femtosecond laser lithography fabrication system, an array of 1960 Mach-Zehnder interferometer (MZI) structures are fabricated on a 4-inch wafer in a continuous fabrication process, as shown in Fig. 3(c). We chose the MZIs as an example for testing our system because MZIs are the heart of high-speed EO modulators based on TFLN. It is confirmed that each MZI of 10 mm total length can be patterned within only 4.3 second, indicating a high manufacturing efficiency which is highly desirable for industrial-scale mass-production.

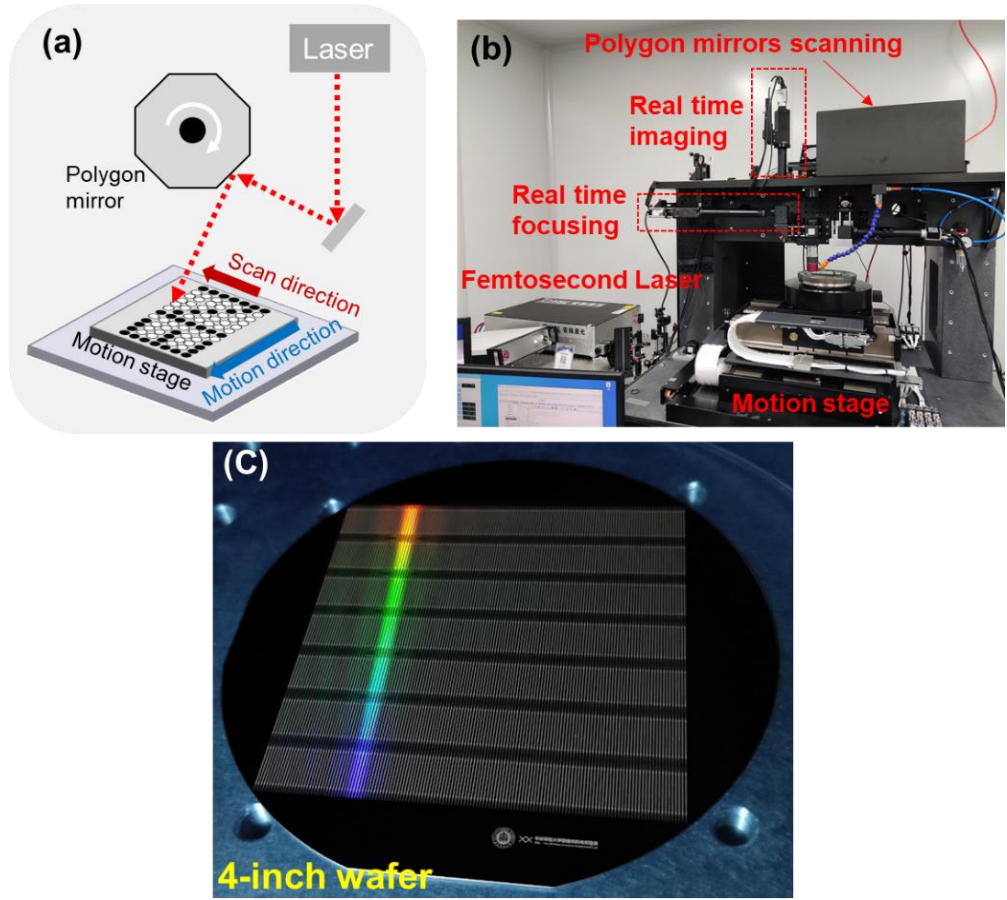

Figure 3. (a) Principle of ultra-high-speed polygon laser scanner. (b) Physical drawing of femtosecond laser lithography system based on polygon scanner. (c) 1960 MZI structures in 4-inch wafer.

## 4. APPLICATIONS OF WAFER-SCALE NANOFABRICATION

### 4.1 Rapid fabrication of large photonic structures on TFLN

Using the high-speed femtosecond laser lithography system, we successfully fabricate photonic structures of large footprint and reasonable propagation loss. In particular, the propagation loss of the waveguide, which is of vital importance for PIC application, should be characterized of sufficient accuracy. For this purpose, we fabricate an unbalanced MZI on the LNOI substrate using the high-speed PLACE fabrication system. The difference in the lengths of the two arms is 10 cm as shown in Fig. 4(a). The waveguide propagation loss can be determined by measuring the extinction ratio (ER) of the transmitted interference signal as a function of the wavelength recorded by the power meter, because two 1×2 multi-mode interferometers (MMI) of 50:50 beam splitting ratio are used near the two ends of MZI. The insets of Fig. 4(a) show the zoom-in optical micrographs of 1×2 MMI and the Euler bends of the wrapped waveguide, as indicated by the arrows. We used a tunable laser covering the wavelength range of telecom band as the light source for measuring the waveguide loss. The tunable laser was coupled into the LN ridge waveguide using a lensed fiber. The output signal from the LN waveguide was collected by another lensed fiber and directed into the photodetector. The electric signals converted by the exiting light from the MZI were analyzed by an oscilloscope. The transmission spectrum near the 1563.84 nm is plotted by Fig. 4(b), showing the free spectrum range (FSR) of 0.0116 nm and an (ER of ~-5 dB. The optical path difference can be expressed as $\frac{\lambda^2}{FSR}$ of 21 cm which agrees well with the 10 cm length difference in the two arms of the unbalanced MZI multiplied by the group refractive index of 2.1 in the LN ridge waveguide. Given that the output powers of the short and long arms are P1 and P2, respectively, and the length difference between the two arms is L, then the ER can be expressed as $ER = \frac{P1-P2}{P1+P2}$, indicating the propagation loss $\alpha = -\frac{10\log\left(\frac{P2}{P1}\right)}{L} = -10\log(\frac{[1-ER]}{[1+ER]})/L$. Based on the experimentally measured ER of -5 dB, we derive the propagation loss of 0.3 dB/cm for the fabricated waveguides. The loss is approximately one order of magnitude higher than the typical waveguide loss in the LN waveguides fabricated using PLACE technique, because previously a different scan scheme of the focused femtosecond laser beam was adopted in which the femtosecond laser spot was translated along the contours of the waveguides to ensure generation of smooth edges of the Cr hard mask[19,20]. Alternatively, in the current high-speed femtosecond laser lithography fabrication, a raster scan scheme is used which leads to higher fabrication efficiency but slightly higher edge roughness. Nevertheless, the propagation loss of 0.3 dB/cm is comparable to the typical loss in the LN ridge waveguides fabricated by electron beam lithography (EBL), and is already sufficient low for the LN photonic devices of centi-meter scale footprint. Large-scale PICs usually demand lower propagation loss which can be achieved by further shortening the laser switching on-off time and improving the synchronization between the polygon scanner and the high-speed shutter, as these are the major factors responsible for the edge roughness left behind on the Cr mask.

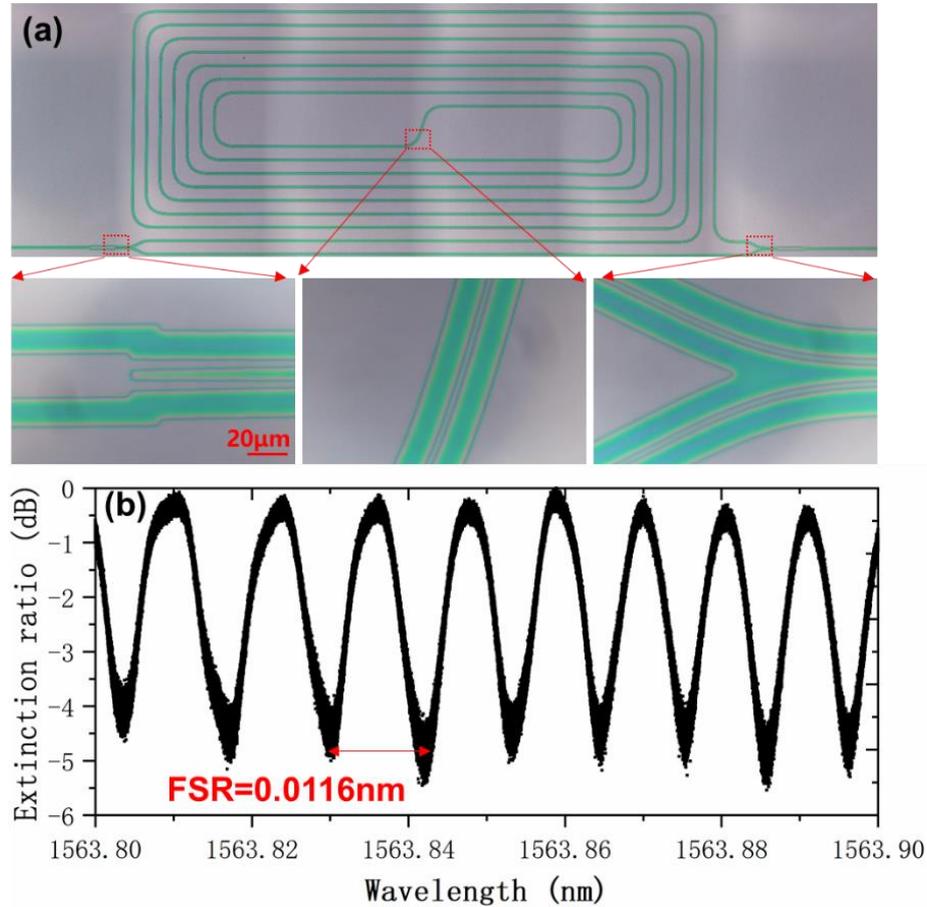

Figure 4. (a) Micrograph of unbalanced MZI with 10 cm length difference on LNOI. Inset: zoom-in micrographs of 1×2 MMI and Euler bend waveguide. (b) The transmission spectra near the 1563.84 nm for indicating the extinction ratio.

## 4.2 Rapid fabrication of wafer-scale optical phase plate

The high-speed femtosecond laser lithography system is a valuable fabrication tool for applications requiring large fabrication areas and sub-micron fabrication resolutions on various kinds of materials, which is not limited for TFLN-based PIC application. Below, we demonstrate fabrication of wafer-scale OPPs, which are fundamental to light field shaping and tailoring. The first wafer-scale OPP with a size of 1 cm × 1 cm is composed of a checkerboard grating consisting of 1-μm units as shown in Fig. 5(a)-(b). The digital camera picture of the OPP presented in Fig. 5(b), we can see the structural color due to the optical diffraction. To check its diffraction performance, a continuous wave (CW) laser at the wavelength of 632 nm propagates through the OPP, and then recorded by the CCD camera to reveal the diffraction pattern. As shown in Fig. 5(c), we can see that the CW beam was split into for diffracted beams at the corner of a square. In the middle of the image, one can see the un-diffracted zeroth-order beam. The uniformity in the four diffracted beams, in terms of both the beam intensity and its spatial profile, clearly indicates the uniformity of the fabricated OPP structure. Similarly, a vortex grating as illustrated in Fig. 5(d)-(e) was fabricated using the same fabrication procedures. The 632-nm CW laser beam was split into eight beams equally spaced in an annular ring, as shown in Fig. 5(f).

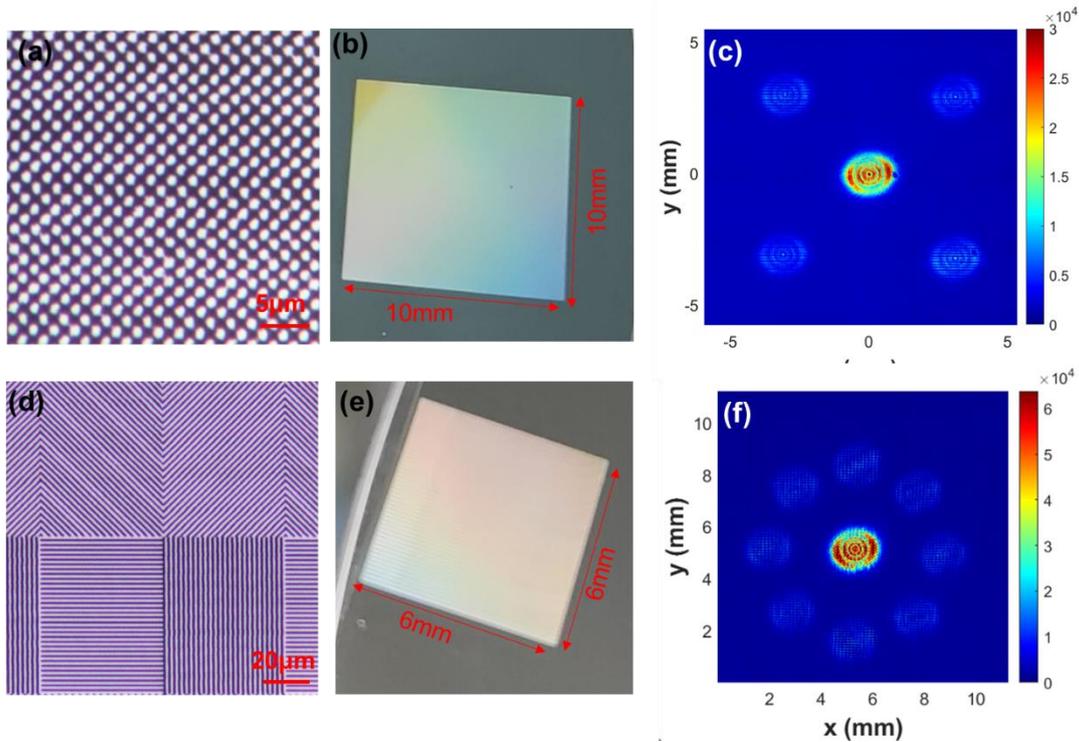

Figure 5. (a) Micrograph of checkerboard grating with 1 μm unit pixel. (b) The digital camera photograph of Fig. 6(a) with the size of 10 mm×10 mm. (c) Spot distribution after the checkerboard grating. (d) Micrograph of vortex grating with 1 μm linewidth. (b) The digital camera photograph of Fig. 6(d) with the size of 6 mm×6 mm. (f) Spot distribution after the vortex grating.

## 4.3 Rapid wafer-scale color printing

The capability of rapidly producing the fine grating structures with micro-scale pixel size and wafer-scale footprint allows for wafer-scale color printing based on the grating diffraction effect. The structural colors generated this way strongly depends on the diffraction angle and the grating period, as shown in Fig. 6(a). One can generate the full range of spectrum in the visible by tuning the grating period for a fixed angle of view (i.e., the fixed diffraction angle). Here, we chose to produce reflective gratings on the fused silica wafer coated with Cr thin film, which is similar to the grating structures described in Sect. 4.2. However, in the color printing, the grating in each colorful pixel of a size of 20 μm ×20 μm is fabricated with its own grating period corresponding to the color borne by the pixel. Here, the angle of view, which is the same as the diffraction angle of our color printing sample, is designed to be 30 degrees, which is a convenient angle for the viewer holding the sample in his/her hand. According to such a diffraction angle, we can calculate the grating period d as a function of the diffraction wavelength λ using the well-known equation $d\sin\theta = m\lambda$. The calculated result is presented in Fig. 6(b) by choosing a diffraction order m=4, so as to have a relatively large grating period (i.e., from 3.2-6.2 μm) for the visible wavelengths. To confirm the calculation result in Fig. 6(b), we fabricated a group of samples of different colors (i.e., the color palettes as being called hereafter) for the characterization of the quality of our color printing. As shown in Fig. 6(c), the digital-camera-captured photos of the fabricated color palettes under the illumination of a white light source show exactly the same color as presented in Fig. 6(b) for the corresponding grating periods. As demonstrated in Fig. 6(d), the reflected beams from the palettes produce a range of colors to form a wide gamut in the CIE 1931 chromaticity diagram, which means that the proposed grating diffraction effect can be used for wafer-scale color printing application thanks to the high fabrication efficiency and high fabrication resolution of our polygon-scanner-based femtosecond laser lithography system. Figures 6(e)-(f) provide the comparison between an original hard-copy picture of the campus of East China Normal University (Fig. 6(e)) and the white balanced digital photograph of the same image printed on the 4-inch wafer using our femtosecond laser lithography system (Fig. 6(f)). It can be seen that the original image is well reproduced in the color printing sample despite that some difference does exist in the two images which can be improved by further refining the grating design parameters and image conversion algorithm. The total time for printing out the full picture of an area of 32.7 cm2 as shown in Fig. 6(f) is about ~6.8 hrs, thanks to the ultra-high fabrication efficiency of 4.8 cm$^2$/h.

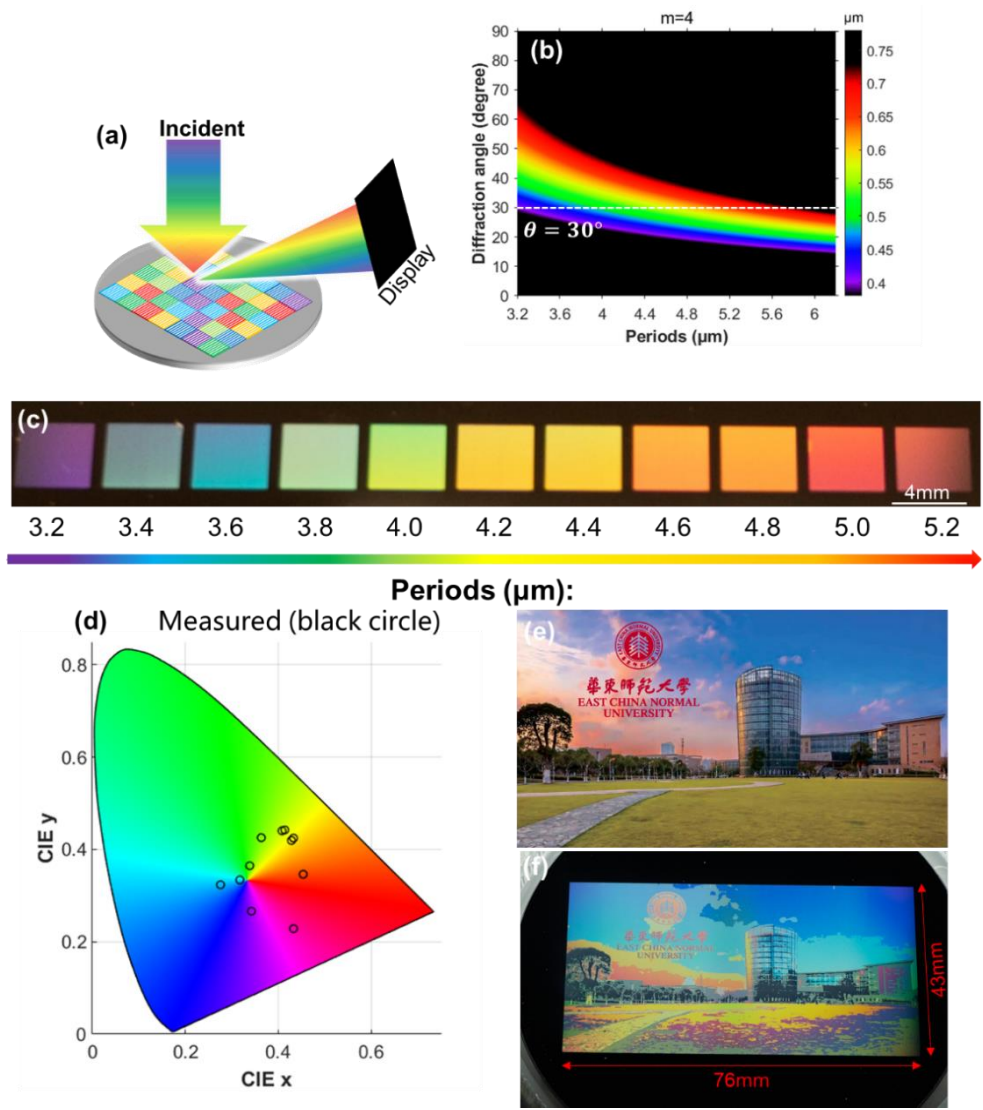

Figure 6. (a) The principle of structural colors based on grating diffraction effect. (b) Theoretically calculated diffraction angle as a function of grating periods in the 4-order diffraction. (c) The digital camera photograph of the color palettes at different grating periods. (d) The measured reflection colors (black circles) are mapped in the CIE 1931 chromaticity diagram. Campus of East China Normal University of (e) original picture and (f) 4-inch wafer color printing.

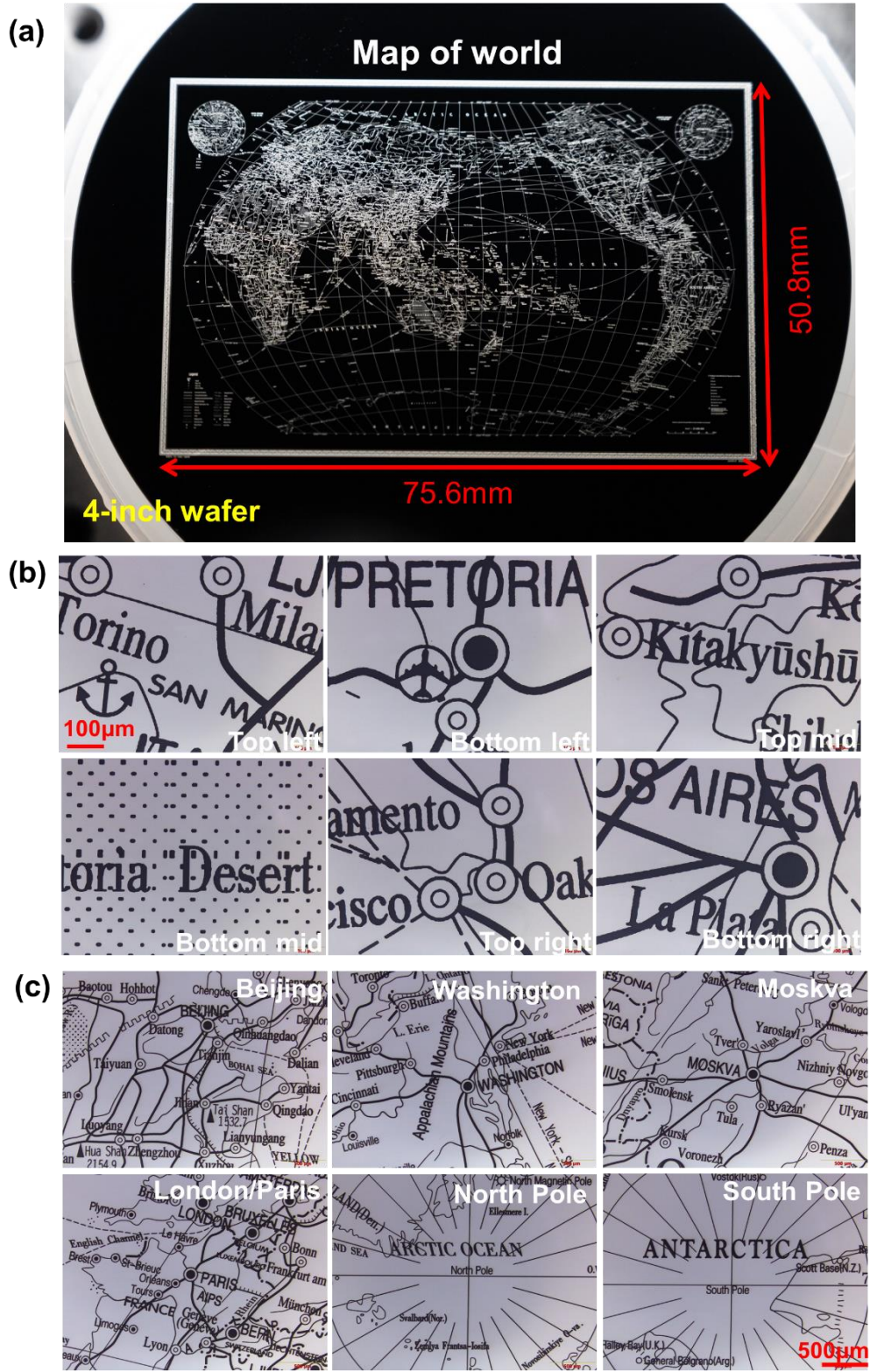

Figure 7 (a) The digital camera photograph of the map of world. (b) Micrographs at the positions of top left, bottom left, top mid, bottom mid, top right, and bottom right of world map. (c) Micrographs of map in different cities (Beijing, Washington, Moskva, London and Paris) and North/South Pole for global view.

## 4.4 Wafer-scale micro-printing of ultra-high resolution and high uniformity

The biggest advantage of our polygon-scanner based femtosecond laser lithography system is the essential IFOV and extremely high uniformity across the whole wafer area. The reason is that with such a strategy, the fabrication resolution is only determined by the size of the focal spot generated by the focal system (i.e., the objective lens). Meanwhile, the fabrication area is only determined by the range of motion of the high-precision stage, which in principle can be extended to meter-scale range with today's commercial product. Thus, the dilemma of creating an optical imaging system which can ensure a high and uniform spatial resolution in a large field of view can be solved. To showcase such a capability of mask patterning with high resolution and high uniformity with IFOV, we carry out the continuous printing of a world map in the 4-inch fused silica wafer coated with Cr thin film. The total area of the world map, as shown in Fig. 7(a), is of a size of 75.6 mm×50.8 mm. Remarkably, the map consists of 86.4 billion pixels in total. To verify the processing uniformity, we indivisually examine the printing quality in various areas of the top left, bottom left, top middle, bottom middle, top right and bottom right of the map, respectively, as presented in Fig. 7(b). The sharp edges of the drawings produced by femtosecond laser direct writing as well as the straightness of the lines in all the panels of Fig. 7(b) clearly indicate the high uniformity of our photolithography system. Figure 7(c) shows the micrographs of different cities including Beijing, Washington, Moskva, London/Paris and South/North Poles featured with intricate traffic networks, again confirming the capability of generating arbitrary patterns with our polygon-scanner based femtosecond laser lithography system.

## 5. CONCLUSIONS

Photolithography is an important micro- and nanofabrication technology for producing miniaturized devices and systems in various applications ranging from electronics and photonics to chemistry, biology and medicine. Photolithography based on femtosecond laser direct writing shows the advantages of high spatial resolution, high fabrication efficiency as compared to the EBL and focused ion beam writing, as well as flexibility in terms of the substrate materials that can be processed with femtosecond laser ablation or femtosecond laser modification. For industrial applications, the high throughput is of vital importance which requires the realization of high scan speed with the photolithography system. Our fabrication system based on the polygon scanner provides an example to accomplish this task. Using such a system, we show that we can fabricate low-loss waveguides on TFLN substrate for large-scale PIC application. We also demonstrate a range of wafer-scale microfabrication applications such as OPP and laser printing. Currently we have achieved a fabrication efficiency of 4.8 cm$^2$/hr at the highest fabrication resolution of 200 nm, which allows to generate Tera-pixel scale patterns within an area of full-sized 4-inch wafer in a few hours. It is expected that with the continuous advances in high-repetition-rate femtosecond laser and high-speed electronic shutter/controller, the fabrication efficiency can be further promoted by 1~2 orders of magnitude, eventually allowing for patterning a 4-inch wafer in a few minutes. This will have profound implication as miniaturization will play the central role in the future society.

## ACKNOWLEDGEMENTS

The research was supported by Science and Technology Commission of Shanghai Municipality (NO.21DZ1101500) and Shanghai Municipal Science and Technology Major Project (Grant No.2019SHZDZX01).